\documentclass[10pt,conference]{IEEEtran}
\usepackage{amsmath}
\usepackage{cite}
\usepackage{graphicx}
\usepackage[table,xcdraw]{xcolor}
\usepackage{bm}
\usepackage{array}
\usepackage{multirow}
\usepackage{makecell}
\usepackage{diagbox}
\usepackage{tabularx}
\usepackage{booktabs}
\usepackage{amssymb}
\usepackage{url}

\begin{document}

\title{A Novel Domain-Aware CNN Architecture \\for Faster-than-Nyquist Signaling Detection}

\author{
Osman Tokluoglu\IEEEauthorrefmark{1}, Enver Cavus\IEEEauthorrefmark{1}, Ebrahim Bedeer\IEEEauthorrefmark{2}, Halim Yanikomeroglu\IEEEauthorrefmark{3}\\
\IEEEauthorrefmark{1}Department of Electrical and Electronics Engineering, Ankara Yildirim Beyazit University, Ankara, Turkiye\\
\IEEEauthorrefmark{2}Department of Electrical and Computer Engineering, University of Saskatchewan, Saskatoon, SK, Canada\\
\IEEEauthorrefmark{3}Department of System and Computer Engineering, Carleton University, Ottawa, ON, Canada\\
\{otokluoglu, ecavus\}@aybu.edu.tr, e.bedeer@usask.ca, halim@sce.carleton.ca
}

\maketitle

\begin{abstract}
This paper proposes a convolutional neural network (CNN)-based detector for faster-than-Nyquist (FTN) signaling that employs structured fixed kernel layers with domain-informed masking to mitigate intersymbol interference (ISI). Unlike standard CNNs with sliding kernels, the proposed method utilizes fixed-position kernels to directly capture ISI effects at varying distances from the central symbol. A hierarchical filter allocation strategy is also introduced, assigning more filters to earlier layers for strong ISI patterns and fewer to later layers for weaker ones. This design improves detection accuracy while reducing redundant operations. Simulation results show that the detector achieves near-optimal bit error rate (BER) performance for $\tau \geq 0.7$, closely matching the BCJR algorithm, and offers computational gains of up to $46\%$ and $84\%$ over M-BCJR for BPSK and QPSK, respectively. Comparative analysis with other methods further highlights the efficiency and effectiveness of the proposed approach. To the best of our knowledge, this is the first application of a fixed-kernel CNN architecture tailored for FTN detection in the literature.
\end{abstract}


\section{Introduction}
\label{sec:introduction}
The growing demand for high data rates and limited spectral resources has spurred the development of techniques aimed at improving spectral efficiency. One such technique is faster-than-Nyquist (FTN) signaling, which increases transmission rates by deliberately introducing intersymbol interference (ISI) \cite{mazo1975}. Although optimal algorithms like BCJR handle the resulting ISI effectively, their computational demands are high \cite{bahl1974}, prompting the need for low-complexity detection methods with strong performance.

Several reduced-complexity approaches have been proposed. Prlja and Anderson introduced the M-BCJR algorithm with minimum-phase preprocessing to reduce complexity while maintaining bit error rate (BER) performance across various compression factors ($\tau$) \cite{prlja2012}. However, its complexity still scales exponentially with modulation order. Sugiura et al. proposed a frequency-domain equalization (FDE) method using cyclic prefixes and minimum mean square error (MMSE) detection, achieving good performance at roll-off factor $\beta = 0.5$, but at the cost of increased bandwidth and reduced spectral efficiency \cite{sugiura2013}. A symbol-by-symbol estimation strategy with a go-back-$K$ refinement was proposed in \cite{bedeer2017}, offering low complexity but limited performance at lower $\tau$ values such as $\tau = 0.7$. Similarly, the simplified M-BCJR detector based on Ungerboeck’s model in \cite{li2018} improves performance for BPSK at $\tau = 0.8$ but becomes impractical for higher modulation orders.

In parallel, deep learning has gained traction in FTN detection. Most studies combine neural networks with traditional methods, while only a few explore deep learning as standalone detectors. For instance, detectors based on Long Short-Term Memory (LSTM) networks in \cite{baek2021} match BCJR performance for $\tau = 0.8$ with lower complexity. Deep learning-aided uccessive interference cancellation (SIC) and end-to-end receivers were proposed in \cite{song2020}, showing robustness for $\tau = 0.8$. In \cite{pan2020}, a fully connected deep neural network (FC-DNN) with a sliding-window design was used for FTN signaling combined with non-orthogonal multiple access (FTN-NOMA) in internet of Things (IoT), outperforming MMSE-FDE. The  deep learning-enhanced list sphere decoding (DL-LSD) algorithm in \cite{abbasi2022} combines recurrent neural networks (RNNs) and dense layers for efficient list sphere decoding. Bi-LSTM networks were used in \cite{baek2023} to eliminate the need for explicit equalization in multipath fading scenarios. The deep learning-assisted sum-product detection algorithm (DL-SPDA) model in \cite{liu2021} incorporates neural components into sum-product decoding, improving ISI mitigation at $\tau = 0.5$ and $\tau = 0.6$.

While RNNs such as LSTM and Bi-LSTM are effective, they come with high training and inference complexity. In contrast, convolutional neural networks (CNNs) offer parallelism and efficient feature extraction. Although underexplored in FTN detection, 1D CNNs have proven effective in signal processing tasks like orthogonal frequency division multiplexing (OFDM) detection \cite{kiranyaz2020review, WANG2023102055}, and have been used in hybrid CNN-BiLSTM designs for FTN signaling \cite{yang2024mhsa}. De Filippo et al. recently applied skip-connected CNNs to FTN detection \cite{defilippo2025}, achieving promising results. However, their use of standard convolutions without domain-specific ISI modeling may lead to inefficiencies.

Motivated by this gap, we propose a CNN-based FTN detector that employs structured fixed kernel layers with domain-informed masking to explicitly learn ISI patterns at varying positions. A hierarchical filter strategy assigns more filters to early layers for stronger ISI and fewer to later layers. This structured design improves feature learning, reduces redundancy, and enhances accuracy while maintaining computational efficiency. To the best of our knowledge, this is the first application of fixed-kernel CNNs tailored for FTN detection. Simulation results show that our model matches BCJR-level BER performance for $\tau \geq 0.7$, while achieving up to $46\%$ and $84\%$ reductions in computational complexity over M-BCJR for BPSK and QPSK, respectively.

The remainder of the paper is structured as follows: Section~\ref{sec:system_model} describes the FTN system model. Section~\ref{sec:cnn_tech} presents the proposed CNN-based detector. Section~\ref{sec:Simulation Results} discusses simulation results, and Section~\ref{sec:conclusion} concludes the paper.

\section{System Model}
\label{sec:system_model}
For a baseband pulse \(g(t)\) with bandwidth \((1 + \beta)/2T\), where \(\beta\) is the roll-off factor, the transmitted signal is given as
\begin{equation}
s(t) = \sum_{k} a_k g(t - k{\tau}T),
\label{eq:transmission_signal}
\end{equation}
where \(a_k\) is the $k$-th modulated symbol (BPSK or QPSK), and \(T\) denotes the Nyquist interval. The compression factor \(\tau \in (0, 1)\) controls the symbol rate relative to the Nyquist rate. Reducing \(\tau\) increases spectral efficiency by accelerating symbol placement, but introduces ISI due to pulse overlap, as the pulse bandwidth \((1 + \beta)/2T\) becomes insufficient for the tighter spacing. Accurate detection thus requires compensation for the resulting ISI. The pulse \(g(t)\) is normalized to unit energy as
\begin{equation}
\int_{-\infty}^{+\infty} |g(t)|^2 \, dt = 1.
\end{equation}

A root-raised cosine (RRC) pulse with \(\beta = 0.35\) is typically used \cite{benedetto1993}. The spectral efficiency gain is approximately \(1/\tau\); for instance, \(\tau = 0.9\) yields an 11\% gain.

Transmission is considered over an AWGN channel with noise distributed as \(\mathcal{N}(0, N_0 / 2)\). After filtering, the received signal becomes
\begin{equation}
y(n\tau T) = \sum_{k} a_kx((n - k)\tau T) + w(n\tau T), 
\label{eq:received_signal}
\end{equation}
where \(x(t)\) is the convolution of the transmit filter with its time-reversed version:
\begin{equation}
x(t) = g(t) * g(-t),
\end{equation}
and \(w(n\tau T)\) denotes colored noise at the sampling instant.

\begin{figure}[!t]
\centering
\includegraphics[width=\linewidth]{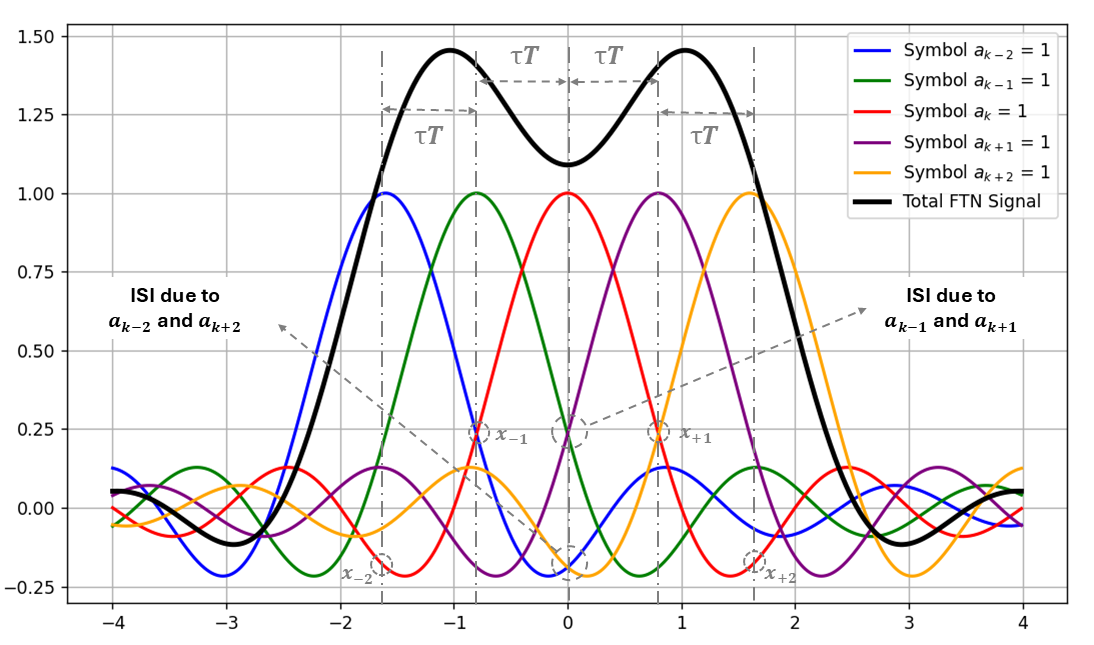}
\caption{Faster-than-Nyquist signaling structure and interference coefficients.}
\label{fig:Fig1_FTN_Structure}
\end{figure}

Fig.~\ref{fig:Fig1_FTN_Structure} illustrates FTN signaling for a five-symbol BPSK sequence 
$
\begin{bmatrix}
1 & 1 & 1 & 1 & 1
\end{bmatrix}
$,
where detection of \(a_k\) is influenced by its adjacent symbols. The ISI span depends on \(\tau\), with the one-sided ISI length denoted by \(N\). In the example shown, \(N = 2\), and the black curve represents the composite signal, while colored curves show contributions of neighboring symbols. The ISI filter coefficients for \(\tau = 0.8\) are 
$
\begin{bmatrix}
x_{-2} & x_{-1} & x_{+1} & x_{+2}
\end{bmatrix}
$,
where \(x_{\pm1}\) and \(x_{\pm2}\) are first- and second-order coefficients, respectively. The received sample \(y_k\) can be expressed as
\begin{equation}
\scalebox{0.85}{$
y_k = 
\big[ x_N \, x_{N-1} \, \dots \, x_1 \, x_0 \, x_1 \, \dots \, x_{N-1} \, x_N \big]
\begin{bmatrix}
a_{k-N} \\ 
a_{k-N+1} \\ 
\vdots \\ 
a_{k-1} \\ 
a_k \\ 
a_{k+1} \\ 
\vdots \\ 
a_{k+N-1} \\ 
a_{k+N}
\end{bmatrix}
+ w_k
.$}
\label{eq:received_signal}
\end{equation}

For a length-$K$ sequence and \(N = 2\), the complete system can be written in matrix form as
\vspace{-0.5cm}
\begin{multline}
\\
\begin{bmatrix}
y_0 \\ 
y_1 \\ 
y_2 \\ 
y_3 \\ 
y_4 \\ 
y_5 \\ 
y_6
\end{bmatrix}
=
\begin{bmatrix}
x_0 & x_1 & x_2 & 0 & 0 & 0 & 0 \\ 
x_1 & x_0 & x_1 & x_2 & 0 & 0 & 0 \\ 
x_2 & x_1 & x_0 & x_1 & x_2 & 0 & 0 \\ 
0 & x_2 & x_1 & x_0 & x_1 & x_2 & 0 \\ 
0 & 0 & x_2 & x_1 & x_0 & x_1 & x_2 \\ 
0 & 0 & 0 & x_2 & x_1 & x_0 & x_1 \\ 
0 & 0 & 0 & 0 & x_2 & x_1 & x_0
\end{bmatrix}
\begin{bmatrix}
a_0 \\ 
a_1 \\ 
a_2 \\ 
a_3 \\ 
a_4 \\ 
a_5 \\ 
a_6
\end{bmatrix}
+
\begin{bmatrix}
w_0 \\ 
w_1 \\ 
w_2 \\ 
w_3 \\ 
w_4 \\ 
w_5 \\ 
w_6
\end{bmatrix}
.
\label{eq:matrix_full}
\end{multline}

In general, this can be written compactly as follows:
\begin{multline}
\mathbf{y} = \mathbf{X} \mathbf{a} + \mathbf{w}, \quad
\\
\scriptsize
\hspace*{-29.3em}
\begin{bmatrix}
y_0 \\ 
\vdots \\ 
y_{K-1}
\end{bmatrix}
=
\\ 
\scriptsize
\begin{bmatrix}
x_0 & x_1 & x_2 & \cdots & x_N & 0 & \cdots & 0 \\
x_1 & x_0 & x_1 & \cdots & x_{N-1} & x_N & \cdots & 0 \\
x_2 & x_1 & x_0 & \cdots & x_{N-2} & x_{N-1} & \cdots & 0 \\
\vdots & \vdots & \vdots & \ddots & \vdots & \vdots & \ddots & \vdots \\
x_N & x_{N-1} & x_{N-2} & \cdots & x_0 & x_1 & \cdots & x_N \\
0 & x_N & x_{N-1} & \cdots & x_1 & x_0 & \cdots & x_{N-1} \\
0 & 0 & x_N & \cdots & x_2 & x_1 & \cdots & x_{N-2} \\
\vdots & \vdots & \vdots & \ddots & \vdots & \vdots & \ddots & \vdots \\
0 & 0 & 0 & \cdots & x_N & x_{N-1} & \cdots & x_0
\end{bmatrix}
\begin{bmatrix}
a_0 \\
a_1 \\
a_2 \\
a_3 \\
\vdots \\
\\
a_{K-4} \\
a_{K-3} \\
a_{K-2} \\
a_{K-1}
\end{bmatrix}
\\
+
\scriptsize
\begin{bmatrix}
w_0 \\ 
\vdots \\ 
w_{K-1}
\end{bmatrix},
\label{eq:matrix_form}
\end{multline}
where $\mathbf{y}$ is the received vector, $\mathbf{a}$ is the $K \times 1$ data symbol vector, $\mathbf{w} \sim \mathcal{N}(0, \mathbf{X}N_0/2)$ is the noise vector, and $\mathbf{X}$ is the $K \times K$ ISI matrix.

\begin{table}[h!]
\caption{One-sided ISI coefficients for $\tau = 0.7$, $0.8$, and $0.9$.}
\centering
\renewcommand{\arraystretch}{1.5} 
\setlength{\tabcolsep}{6pt} 
\begin{tabular}{>{\centering\arraybackslash}m{1.3cm} c c c}
\rowcolor[gray]{0.9}
\textbf{Coefficient} & \textbf{$\tau = 0.7$} & \textbf{$\tau = 0.8$} & \textbf{$\tau = 0.9$} \\ 
$x_0$ & $9.99 \times 10^{-1}$ & $9.99 \times 10^{-1}$ & $9.99 \times 10^{-1}$ \\ 
\rowcolor[gray]{0.95}
$x_1$ & $3.53 \times 10^{-1}$ & $2.22 \times 10^{-1}$ & $1.02 \times 10^{-1}$ \\ 
$x_2$ & $-1.83 \times 10^{-1}$ & $-1.52 \times 10^{-1}$ & $-7.86 \times 10^{-2}$ \\ 
\rowcolor[gray]{0.95}
$x_3$ & $3.24 \times 10^{-2}$ & $7.62 \times 10^{-2}$ & $4.87 \times 10^{-2}$ \\ 
$x_4$ & $3.16 \times 10^{-2}$ & $-2.44 \times 10^{-2}$ & $-2.26 \times 10^{-2}$ \\ 
\rowcolor[gray]{0.95}
$x_5$ & $-2.79 \times 10^{-2}$ & $5.93 \times 10^{-3}$ & $1.24 \times 10^{-2}$ \\ 
$x_6$ & $1.37 \times 10^{-2}$ & $3.16 \times 10^{-3}$ & $-3.61 \times 10^{-3}$ \\ 
\rowcolor[gray]{0.95}
$x_7$ & $3.31 \times 10^{-4}$ & $-1.73 \times 10^{-3}$ & $9.81 \times 10^{-4}$ \\ 
$x_8$ & $-1.73 \times 10^{-3}$ & $6.64 \times 10^{-4}$ & $-1.69 \times 10^{-4}$ \\ 
\end{tabular}
\label{tab:isi_coefficients}
\end{table}

Table~\ref{tab:isi_coefficients} presents ISI coefficients from $x_0$ to $x_8$ for $\tau = 0.7$, $0.8$, and $0.9$, showing that lower $\tau$ values lead to more significant ISI effects due to increased temporal overlap among symbols.

\section{CNN-Based FTN Signaling Detection Technique}
\label{sec:cnn_tech}

\begin{figure}[!t]
\centering
\includegraphics[width=\linewidth]{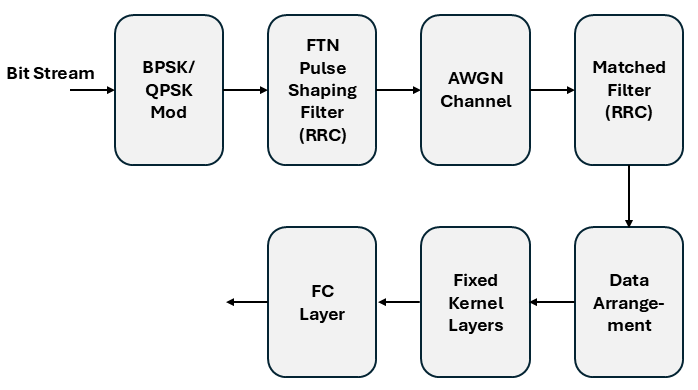}
\caption{FTN signaling-based communication system with proposed signal detector.}
\label{fig:Fig2_Com_Sys_Arch}
\end{figure}

This section introduces the proposed CNN-based FTN detector, designed to address the challenges posed by ISI in FTN signaling. The methodology includes dataset generation, data formatting and system modeling. By incorporating structured fixed kernel layers and domain-informed preprocessing, the detector effectively captures ISI patterns across different $\tau$ values, enabling accurate symbol recovery. The overall FTN communication system with the proposed detector is illustrated in Fig.~\ref{fig:Fig2_Com_Sys_Arch}. Further details are provided in the following subsections.

\subsection{Dataset Preparation}
\label{subsec:dataset}

To train the proposed CNN-based FTN detector, datasets were generated across various $\tau$ and SNR values for BPSK and QPSK modulations over an AWGN channel. Each dataset includes input-output pairs specific to a given $\tau$ and SNR, ensuring diversity for better generalization under different ISI and noise conditions.

High SNR samples were prioritized during training to help the model learn ISI patterns more effectively. This allows the network to better capture the structure of interference before being evaluated in noisier environments. Additionally, separate models were trained for each $\tau$ value, allowing the CNN to tailor its weights to each compression factor and enhance detection accuracy while minimizing generalization errors.

\subsection{Data Arrangement}
\label{subsec:data_arrangement}
In FTN signaling, ISI results from both preceding and succeeding symbols, and its extent depends on the compression factor $\tau$. To capture this, the input to the model is constructed using a context window of $2N+1$ received samples—$N$ on each side of the symbol of interest. Each input sample is structured as:
\begin{equation}
\mathbf{Y}_k = 
\begin{bmatrix}
y_{k-N} & y_{k-N+1} & \cdots & y_k & \cdots & y_{k+N-1} & y_{k+N}
\end{bmatrix},
\label{eq:yk_vector}
\end{equation}
where $y_k$ is the received signal associated with the $k$-th transmitted symbol $a_k$. For edge cases (first and last $N$ samples), zero-padding is applied to maintain consistency in input size.

The value of $N$ was empirically chosen based on simulation results and the one-sided ISI coefficients in Table~\ref{tab:isi_coefficients}. Specifically, $N=2$ for $\tau=0.9$, $N=6$ for $\tau=0.8$, and $N=8$ for $\tau=0.7$. While increasing $N$ improves ISI coverage, it also raises computational cost, so a balance was maintained. For instance, Table~\ref{table:ftn_data_layout_N=6} illustrates the data layout used when $N=6$, where each row is input to the CNN-based detection model.

\begin{table}[ht]
\centering
\caption{Data Structure for $N=6$}
\label{table:ftn_data_layout_N=6}
\renewcommand{\arraystretch}{1.3} 
\setlength{\tabcolsep}{2pt} 
\resizebox{\columnwidth}{!}{ 
\begin{tabular}{cccccccccccccccc}
0th input   & 0 & 0 & 0 & 0 & 0 & 0 & \bm{$y_0$} & $y_1$ & $y_2$ & $y_3$ & $y_4$ & $y_5$ & $y_6$ \\ 
\rowcolor[HTML]{EFEFEF} 
1st input   & 0 & 0 & 0 & 0 & 0 & $y_0$ & \bm{$y_1$} & $y_2$ & $y_3$ & $y_4$ & $y_5$ & $y_6$ & $y_7$ \\ 
2nd input   & 0 & 0 & 0 & 0 & $y_0$ & $y_1$ & \bm{$y_2$} & $y_3$ & $y_4$ & $y_5$ & $y_6$ & $y_7$ & $y_8$ \\ 
\rowcolor[HTML]{EFEFEF} 
3rd input   & 0 & 0 & 0 & $y_0$ & $y_1$ & $y_2$ & \bm{$y_3$} & $y_4$ & $y_5$ & $y_6$ & $y_7$ & $y_8$ & $y_9$ \\ 
4th input   & 0 & 0 & $y_0$ & $y_1$ & $y_2$ & $y_3$ & \bm{$y_4$} & $y_5$ & $y_6$ & $y_7$ & $y_8$ & $y_9$ & $y_{10}$ \\ 
\rowcolor[HTML]{EFEFEF} 
5th input   & 0 & $y_0$ & $y_1$ & $y_2$ & $y_3$ & $y_4$ & \bm{$y_5$} & $y_6$ & $y_7$ & $y_8$ & $y_9$ & $y_{10}$ & $y_{11}$ \\ 
6th input   & $y_0$ & $y_1$ & $y_2$ & $y_3$ & $y_4$ & $y_5$ & \bm{$y_6$} & $y_7$ & $y_8$ & $y_9$ & $y_{10}$ & $y_{11}$ & $y_{12}$ \\ 
\rowcolor[HTML]{EFEFEF} 
$\vdots$    & $\vdots$ & $\vdots$ & $\vdots$ & $\vdots$ & $\vdots$ & $\vdots$ & $\vdots$ & $\vdots$ & $\vdots$ & $\vdots$ & $\vdots$ & $\vdots$ & $\vdots$ \\ 
$(k-6)$th input   & $y_{k-12}$ & $y_{k-11}$ & $y_{k-10}$ & $y_{k-9}$ & $y_{k-8}$ & $y_{k-7}$ & \bm{$y_{k-6}$} & $y_{k-5}$ & $y_{k-4}$ & $y_{k-3}$ & $y_{k-2}$ & $y_{k-1}$ & $y_{k}$ \\
\rowcolor[HTML]{EFEFEF} 
$(k-5)$th input   & $y_{k-11}$ & $y_{k-10}$ & $y_{k-9}$ & $y_{k-8}$ & $y_{k-7}$ & $y_{k-6}$ & \bm{$y_{k-5}$} & $y_{k-4}$ & $y_{k-3}$ & $y_{k-2}$ & $y_{k-1}$ & $y_{k}$ & 0 \\
$(k-4)$th input   & $y_{k-10}$ & $y_{k-9}$ & $y_{k-8}$ & $y_{k-7}$ & $y_{k-6}$ & $y_{k-5}$ & \bm{$y_{k-4}$} & $y_{k-3}$ & $y_{k-2}$ & $y_{k-1}$ & $y_{k}$ & 0 & 0 \\
\rowcolor[HTML]{EFEFEF} 
$(k-3)$th input   & $y_{k-9}$ & $y_{k-8}$ & $y_{k-7}$ & $y_{k-6}$ & $y_{k-5}$ & $y_{k-4}$ & \bm{$y_{k-3}$} & $y_{k-2}$ & $y_{k-1}$ & $y_{k}$ & 0 & 0 & 0 \\
$(k-2)$th input   & $y_{k-8}$ & $y_{k-7}$ & $y_{k-6}$ & $y_{k-5}$ & $y_{k-4}$ & $y_{k-3}$ & \bm{$y_{k-2}$} & $y_{k-1}$ & $y_{k}$ & 0 & 0 & 0 & 0 \\
\rowcolor[HTML]{EFEFEF} 
$(k-1)$th input   & $y_{k-7}$ & $y_{k-6}$ & $y_{k-5}$ & $y_{k-4}$ & $y_{k-3}$ & $y_{k-2}$ & \bm{$y_{k-1}$} & $y_{k}$ & 0 & 0 & 0 & 0 & 0 \\
$k$th input   & $y_{k-6}$ & $y_{k-5}$ & $y_{k-4}$ & $y_{k-3}$ & $y_{k-2}$ & $y_{k-1}$ & \bm{$y_{k}$} & 0 & 0 & 0 & 0 & 0 & 0 \\ 
\end{tabular}
}
\end{table}

\subsection{CNN-Based FTN Detector System Model}
\label{subsec:cnn_model}

A primary challenge in CNN-based FTN detection is accurately modeling ISI introduced by reduced symbol spacing. To address this, we first examined conventional convolutional kernel strategies and their ability to extract ISI-related features \cite{lecun1989}, \cite{lecun1995}. These initial experiments helped uncover the limitations of traditional kernelization and motivated the design of a specialized architecture.

\begin{figure}[ht]
    \centering
    \includegraphics[width=\columnwidth]{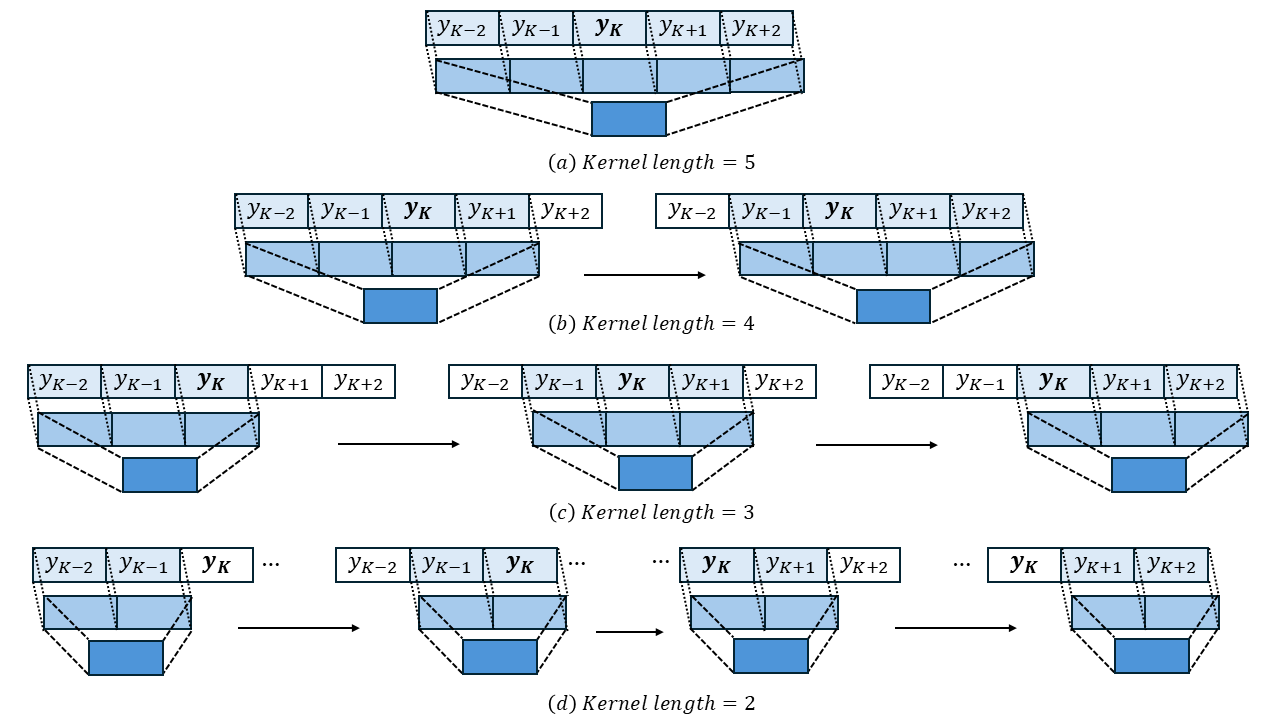} 
    \caption{Illustration of conventional kernelization techniques with varying kernel sizes (\(K = 5\), \(N = 2\)): (a) 5, (b) 4, (c) 3, and (d) 2.}
    \label{fig:Fig3_Conventional_Kernels}
\end{figure}

In conventional CNNs, fixed-size kernels slide across the input to extract features, as shown in Fig.~\ref{fig:Fig3_Conventional_Kernels}. We evaluated this method using sequences of length $K=5$ (with $N=2$ on each side) for $\tau=0.9$. Results showed that large kernels (e.g., size 5) captured irrelevant global patterns, diluting critical ISI interactions. Conversely, overly small kernels (e.g., size 2) lacked sufficient context, struggled to center on the middle symbol, and often misaligned with relevant ISI patterns—compromising both accuracy and generalization.

Among the tested sizes, kernel sizes of 3 and 4 performed best, with size 3 offering an optimal balance between local context and center alignment (see Fig.~\ref{fig:Fig3_Conventional_Kernels}b–c). However, as input sequences grow longer (e.g., $N=6$ for $\tau=0.8$), maintaining this balance becomes difficult, and conventional approaches fall short in capturing extended ISI effects.

\begin{figure}[ht]
    \centering
    \includegraphics[width=\columnwidth]{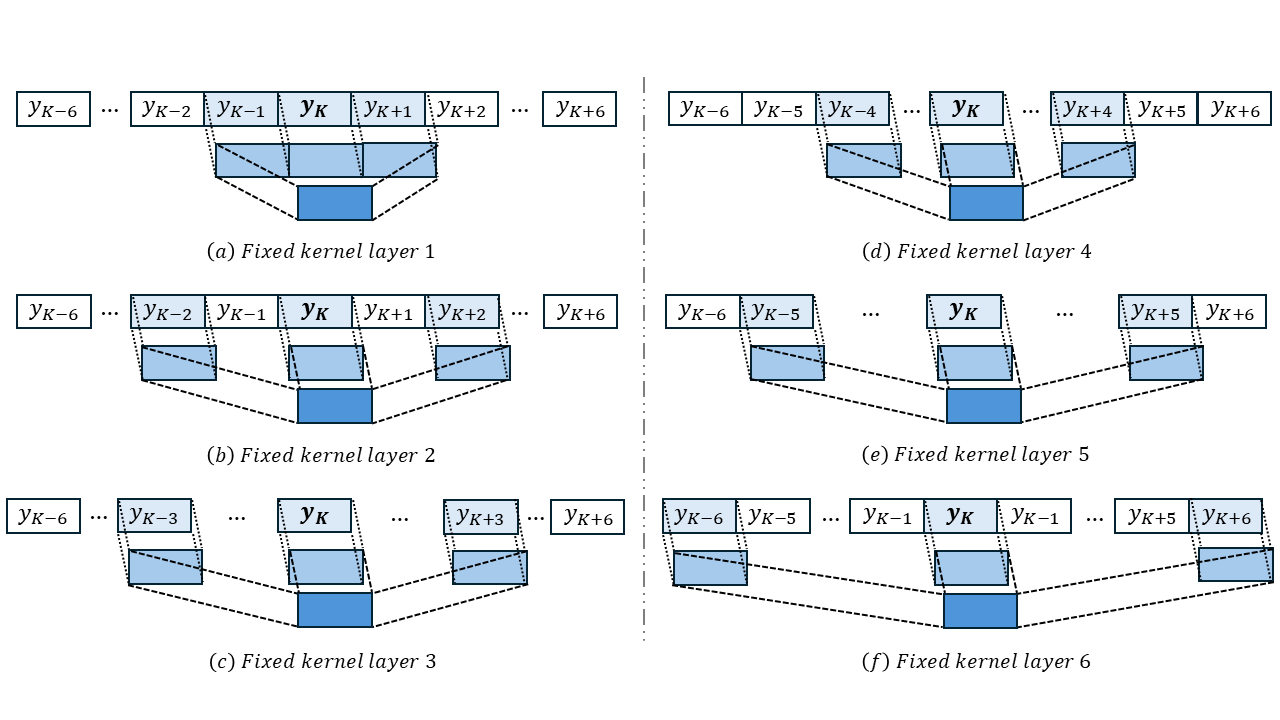} 
    \caption{Illustration of structured fixed kernel layers with domain-informed masking: (a) Layer 1 focuses on symbols one unit away from the middle symbol, (b) Layer 2 captures symbols two units away, (c) Layer 3 focuses on three units away, (d) Layer 4 on four units, (e) Layer 5 on five units, and (f) Layer 6 on six units.}
    \label{fig:Fig4_Fixed_Kernel_Layers}
\end{figure}

To overcome these limitations, we propose a novel method based on \textit{structured fixed kernel layers with domain-informed masking}. As illustrated in Fig.~\ref{fig:Fig4_Fixed_Kernel_Layers}, each layer is dedicated to capturing ISI at a specific distance from the center symbol—layer 1 targets one-unit-away symbols, layer 2 focuses on two-unit-away, and so forth. This structure allows the network to model ISI explicitly and systematically.

To ensure each layer processes only relevant inputs, we apply domain-informed masking. This selectively deactivates unrelated inputs, allowing the kernel to concentrate on its designated ISI context. For instance, with $N=6$, six fixed kernel layers are used, each extracting targeted features to enhance overall detection performance. This structured approach avoids the limitations of sliding kernels and scales effectively with longer sequences. Compared to conventional methods, it improves accuracy and computational efficiency by leveraging explicit ISI modeling.

\begin{figure}[!t]
\centering
\includegraphics[width=\linewidth]{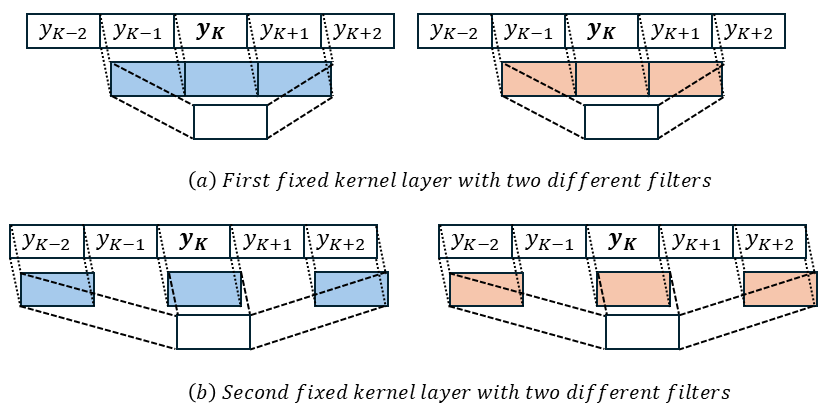}
\caption{Illustration of fixed kernel layers applied with two different filters for \( N=2 \), highlighting diverse feature extraction through varied filter configurations.}
\label{fig:Fig5_Filters}
\end{figure}

\begin{table}[ht]
\centering
\caption{Number of filters for each fixed kernel layer under different $\tau$ values in the proposed CNN detector.}
\renewcommand{\arraystretch}{1.3} 
\begin{tabular}{lccc}
\rowcolor[gray]{0.9}
 & $N=8$ ($\tau$ = 0.7) & $N=6$ ($\tau$ = 0.8) & $N=2$ ($\tau$ = 0.9) \\ 
Layer 1 & 8 & 4 & 2 \\ 
\rowcolor[gray]{0.95}
Layer 2 & 6 & 2 & 1 \\ 
Layer 3 & 4 & 2 & - \\ 
\rowcolor[gray]{0.95}
Layer 4 & 2 & 1 & - \\ 
Layer 5 & 2 & 1 & - \\ 
\rowcolor[gray]{0.95}
Layer 6 & 1 & 1 & - \\ 
Layer 7 & 1 & - & - \\ 
\rowcolor[gray]{0.95}
Layer 8 & 1 & - & - \\ 
\end{tabular}
\label{tab:filter_numbers}
\end{table}

Beyond kernel structure, the number of filters in each fixed kernel layer significantly impacts detection performance. Filters allow the network to extract diverse features from targeted input regions, capturing subtle ISI-related patterns. As shown in Fig.~\ref{fig:Fig5_Filters}, each fixed kernel layer applies multiple filters, with each filter learning different representations of the same ISI region. In the figure, the filters are visually distinguished using blue and orange colors on the kernels, emphasizing their role in capturing diverse features from the input sequence through varied filter configurations. In the illustrated example for $N=2$ (sequence length 5), two fixed kernel layers are used, each employing two filters to enhance feature diversity.

To further optimize this process, we adopt a hierarchical filter allocation strategy. Since ISI is typically stronger near the center symbol, more filters are assigned to earlier layers. For instance, when $N=6$ (as in Fig.~\ref{fig:Fig4_Fixed_Kernel_Layers}, corresponding to $\tau = 0.8$), the first layer uses 4 filters, the next two use 2, and the remaining layers use 1 filter each. This configuration prioritizes critical ISI regions while keeping complexity manageable. The filter settings for various $\tau$ values are summarized in Table~\ref{tab:filter_numbers}.

After the fixed kernel layers, a flatten layer converts the multi-dimensional feature maps into a one-dimensional vector. This output is then passed through a dense layer with four neurons to integrate the extracted features. A final output layer with a single neuron produces the symbol estimate. For QPSK, the real (\( \Re \)) and imaginary (\( \Im \)) parts of the complex symbols are separated and processed independently through identical CNN pipelines.

\section{Simulation Results}
\label{sec:Simulation Results}

\begin{figure}[!t]
\centering
\includegraphics[width=\linewidth]{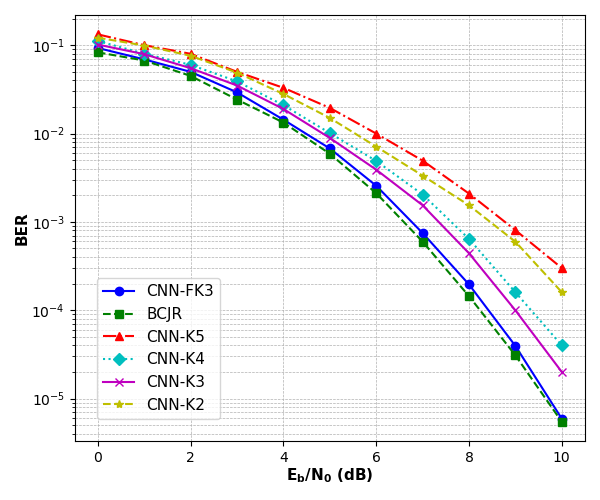}
\caption{Comparison of the BER performance of the proposed CNN-based FTN detector using structured fixed kernel layers (CNN-FK3) with conventional kernel-based CNN architectures (CNN-K2, CNN-K3, CNN-K4, CNN-K5) for $\tau = 0.9$, with a roll-off factor of $\beta = 0.35$.}
\label{fig:Fig6_Fixed_vs_Conventional}
\end{figure}

In order to evaluate the effectiveness of the proposed CNN-based FTN detector with structured fixed kernel layers and domain-informed masking, we first compare it against conventional kernel-based CNNs at $\tau = 0.9$ (sequence length 5, $N = 2$). This setting is chosen because conventional kernels fail to match BCJR-level performance at lower $\tau$ values, where ISI becomes more pronounced. Results are shown in Fig.~\ref{fig:Fig6_Fixed_vs_Conventional}. In the figure, ``CNN-FK3'' refers to our proposed model using a fixed kernel of length 3 (centered on the target symbol and its closest neighbors), as described in Section~\ref{subsec:cnn_model}. ``CNN-K2,'' ``CNN-K3,'' ``CNN-K4,'' and ``CNN-K5'' denote conventional CNNs with kernel lengths 2 to 5, respectively, whose configurations were illustrated in Fig.~\ref{fig:Fig3_Conventional_Kernels}.

\begin{figure}[!t]
\centering
\includegraphics[width=\linewidth]{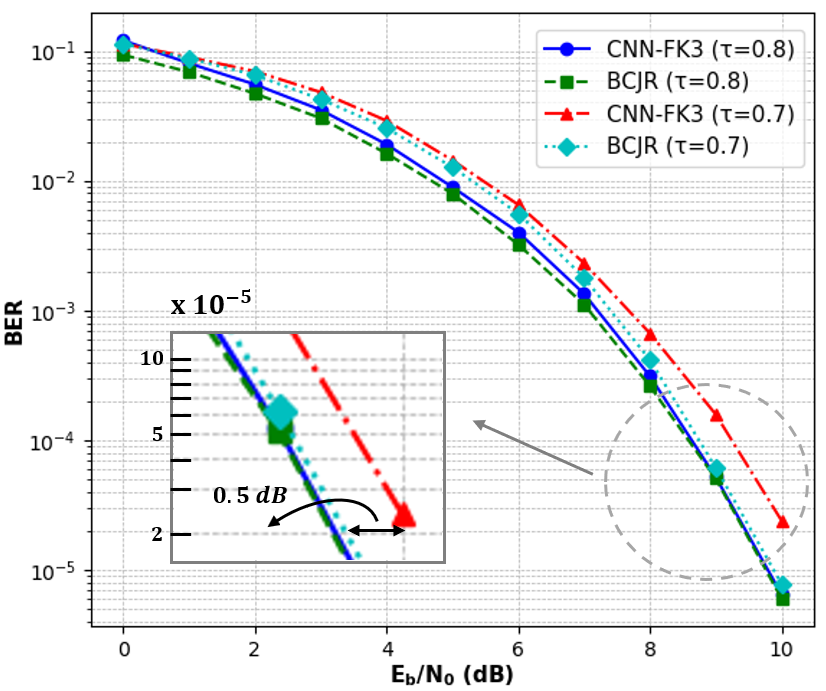}
\caption{Comparison of the BER performance of the proposed CNN-based FTN detector with the optimal BCJR algorithm for $\tau = 0.8$ and $\tau = 0.7$, with a roll-off factor of $\beta = 0.35$.}
\label{fig:Fig7_Cnnfk_vs_Bcjr}
\end{figure}

\begin{figure}[!t]
\centering
\includegraphics[width=\linewidth]{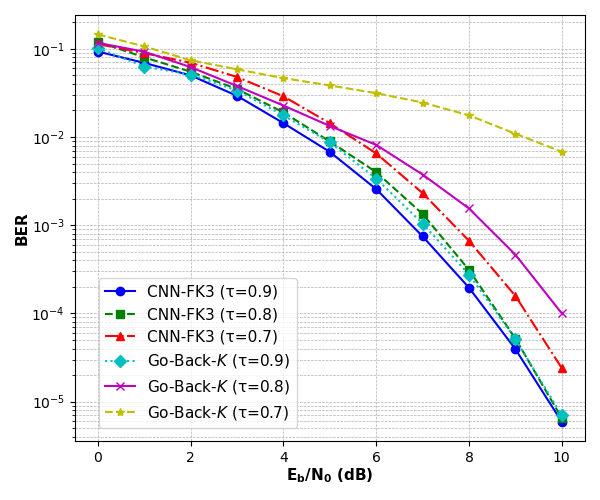}
\caption{BER performance comparison between the proposed CNN-based FTN detector and the Go-Back-\textit{K} algorithm [5] for different compression factors (\(\tau = 0.9\), \(\tau = 0.8\), and \(\tau = 0.7\)), with a roll-off factor of $\beta = 0.35$.}
\label{fig:Fig8_Cnnfk_vs_GobackK}
\end{figure}

As shown in Fig.~\ref{fig:Fig6_Fixed_vs_Conventional}, the proposed CNN-FK3 achieves BER performance nearly identical to the optimal BCJR algorithm for $\tau = 0.9$, clearly outperforming conventional kernel-based models. Among the baselines, CNN-K3 performs best, but still falls short of CNN-FK3. Larger kernels (e.g., size 5 in Fig.~\ref{fig:Fig3_Conventional_Kernels}(a)) fail to capture localized ISI due to their broad receptive fields, while smaller kernels (e.g., size 2 in Fig.~\ref{fig:Fig3_Conventional_Kernels}(d)) lack sufficient context. Mid-sized kernels (sizes 3 and 4) offer better alignment and ISI modeling, as seen in Fig.~\ref{fig:Fig3_Conventional_Kernels}(b–c), with size 3 striking a favorable balance, as discussed in Section~\ref{subsec:cnn_model}. Still, the structured fixed kernel method achieves superior results by precisely targeting ISI at fixed distances and integrating domain knowledge.

To further assess the proposed CNN-based FTN detector, additional experiments were conducted for $\tau = 0.8$ and $\tau = 0.7$, where ISI becomes more severe. The results, shown in Fig.~\ref{fig:Fig7_Cnnfk_vs_Bcjr}, highlight the detector’s robustness under challenging conditions. As BPSK and QPSK yield nearly identical BER performance, a single plot is used for both modulations. At $\tau = 0.8$, the detector closely matches BCJR performance across all SNRs, confirming the effectiveness of the structured fixed kernel approach in mitigating ISI. For $\tau = 0.7$, despite increased ISI, the proposed model remains strong—showing only a minor gap of about 0.5 dB from BCJR at a BER of $2 \times 10^{-5}$. Achieving such performance with a standalone CNN is noteworthy, given that prior works often rely on hybrid models like CNN-BiLSTM \cite{yang2024mhsa}, \cite{liu2022datadriven}. These results demonstrate that fixed kernel CNNs can serve as a competitive and efficient alternative to traditional or hybrid FTN detectors, even under severe ISI conditions.

To examine the trade-off between performance and complexity, we compare the proposed CNN-based FTN detector with the low-complexity Go-Back-\textit{K} successive symbol-by-symbol estimator from \cite{bedeer2017}, using a roll-off factor of $\beta = 0.35$. As both methods prioritize efficiency, this provides a meaningful benchmark. Fig.~\ref{fig:Fig8_Cnnfk_vs_GobackK} shows the BER comparison. At $\tau = 0.9$, both methods yield similar BER performance under mild ISI. However, at $\tau = 0.8$ and $\tau = 0.7$, the proposed detector significantly outperforms the method in \cite{bedeer2017}, demonstrating its ability to handle stronger ISI conditions more effectively.

In order to investigate computational efficiency, the proposed CNN-based FTN detector was compared with the M-BCJR algorithm \cite{kokshoorn2016} and the low-complexity Go-Back-\textit{K} method. The evaluation was based on LUT-weighted hardware cost, which reflects the number of look-up tables (LUTs) required for implementation on FPGA platforms, serving as a practical measure of hardware resource usage. Compared to M-BCJR, the proposed method achieved up to 46\% and 84\% reductions in computational complexity for BPSK and QPSK, respectively. In addition, when benchmarked against the Go-Back-\textit{K} algorithm, the proposed detector demonstrated approximately 39\% lower computational cost. These results underscore the efficiency of the proposed approach, particularly for higher-order modulations.

\section{Conclusion}
\label{sec:conclusion}
In conclusion, this study presents a novel CNN-based FTN detector that employs structured fixed kernel layers with domain-informed masking to explicitly model ISI at varying distances. Unlike conventional CNNs with sliding kernels, the proposed approach uses fixed-position kernels and a hierarchical filter allocation strategy to enhance both detection accuracy and computational efficiency. The detector achieves near-optimal BER performance for $\tau \geq 0.7$, matching BCJR levels while reducing complexity by up to 46\% for BPSK and 84\% for QPSK compared to M-BCJR. Evaluations against existing methods confirm its robustness and efficiency, establishing the proposed architecture as a scalable and low-complexity alternative for FTN detection.

\end{document}